\documentclass{aa}
\usepackage{psfig}
\usepackage{graphicx}

\begin{document}

%
   \title{Discovery of the X-ray burster SAX~J1752.3--3138}

\author{ M. Cocchi \inst{1}, A. Bazzano \inst{1}, L.Natalucci \inst{1}, P. Ubertini \inst{1},
            J. Heise \inst{2}, E. Kuulkers \inst{2,3}, R. Cornelisse \inst{2,3} and J.J.M. in 't Zand \inst{2,3}
          }
   \offprints{M. Cocchi, wood@ias.rm.cnr.it}
 
   \institute{$^{1}$ Istituto di Astrofisica Spaziale ({\em IAS/CNR}), via Fosso del Cavaliere 100, 00133 Roma, Italy \\
              $^{2}$ Space Research Organization Netherlands {\em (SRON)}, Sorbonnelaan 2, NL -- 3584 CA, Utrecht, the Netherlands \\
              $^{3}$ Astronomical Institute, Utrecht University, P.O. Box 80,000, NL -- 3508 TA, Utrecht, the Netherlands
             }

   \date{Received ?; accepted ?}

   \authorrunning{Cocchi et al.}

\abstract{
During a 50 ks monitoring observation of the Galactic bulge performed in September 1999 
by the Wide Field Cameras on board the {\em BeppoSAX} satellite, an X-ray burst was 
detected from a sky position $\sim 3\degr$ off the Galactic centre.
No previously known X-ray sources are located within the position error circle of the 
observed burst.
The new burster, SAX~J1752.3$-$3138, did not show any persistent emission during the
whole observation. No other bursting events, as well as steady emission, were reported 
so far by other instruments or detected in the WFC archive, which covers $\sim 6$ Ms and 
$\sim 4$ Ms for burst and persistent luminosity detection, respectively, starting from 
August 1996.  Unless the source is a very weak transient, this could indicate 
SAX J1752.3$-$3138 is an atypical burster, a member of a possibly new class of sources 
characterised by very low steady luminosities and accretion rates 
($L_{{\rm X}} \la 10^{35} {\rm erg s}^{-1}$) and extremely rare bursting activity.
The characteristics of the detected burst are consistent with a type I event, identifying 
the source as a weakly magnetised neutron star in a low-mass X-ray binary system.
Evidence for photospheric radius expansion due to Eddington-limited burst
luminosity allows to estimate the distance to the source ($\sim 9$ kpc).
}

   \maketitle


\section{Introduction: the {\em BeppoSAX}-WFC Galactic bulge monitoring program}

One of the main scientific goals of the Wide Field Cameras (WFC, \cite{Jage97}) on board the
{\em BeppoSAX} satellite (\cite{Boel97}) is the study of the timing and spectral behaviour of both 
transient and persistent sources in the Galactic bulge, X-ray binaries in particular, 
on time scales ranging from seconds to years.
To this end, a monitoring program of systematic wide field observation of the Sgr~A
sky region is carried out since August 1996 (\cite{Heis98,Heis99,Uber99}).
The program consists of a series of observations, each lasting 60 ks net time, almost weekly spaced 
throughout the two visibility periods (August-October and February-April) of the Galactic Centre 
region. Up to January 2001, an amount of 4.0 Ms of data has been accumulated this way.
The WFC program on the Galactic bulge is significantly contributing in the study of 
transient X-ray binary systems, mainly X-ray bursters (see e.g. \cite{Zand00} for a recent overview).

SAX J1752.3$-$3138 was discovered by the WFCs during the Summer--Fall 1999 monitoring campaign 
of the Galactic bulge.  The source was observed during an X-ray burst (see Fig.1 and Fig. 2, top panel) and no 
persistent emission was found (\cite{Cocc99a}).  No other X-ray bursts are detected in the whole 
WFC data archive. The net exposure time available for burst detection is 6.3 Ms up to January 2001, 
being the attitude constraints not so stringent as for steady emission detection.  This makes 
the observed event the only source of information about SAX J1752.3$-$3138 available so far.
Nevertheless, burst detection itself provides important details on a newly discovered source, 
being such events uniquely associated to weakly magnetised neutron stars (NS) in Low-Mass X-ray Binary 
systems (LMXBs).
Besides the nature of the source, in several cases X-ray bursts allow to estimate useful parameters 
such as the source distance and luminosity, and the NS radius (see Lewin, van Paradijs \& Taam 1993, 
hereafter \cite{Lewi93}, for a comprehensive review).
A single detected burst in a fairly big amount of observing time is even more interesting, since 
its uniqueness puts important constraints on the characteristics of the accretion onto the neutron star.
In the case of SAX J1752.3$-$3138 in particular, since no steady emission was observed, we explore the 
bursting behaviour of sources at very low persistent luminosities. The study of the lower luminosity 
boundary when NS start bursting is an important ingredient for our knowledge of thermonuclear bursts.

In the next section we report on the observation and the spectral/timing 
data analysis of the (so far) only burst of SAX J1752.3$-$3138.
We then briefly discuss the results in Section 3, proposing SAX J1752.3$-$3138 either as a weak 
transient NS-LMXB or, more intriguingly, as a member of a possibly new class of low-luminosity, slowly 
accreting X-ray bursters. An estimate of the distance to the source will be also given.

\begin{table}[ht]
\caption{Summary of the main parameters of the observed burst
}
\protect\label{t:ae}
\begin{flushleft}
\begin{tabular}{lc}
\hline
\hline \noalign{\smallskip}
Burst date                            & 1999, September 2 \\
Burst UT time (h)                     & 3.3936 \\
{\em e}-folding time (2--19 keV)      & $21.9\pm 1.3$ s  \\
{\em e}-folding time (2--6 keV)       & $21.6\pm 2.8$ s  \\
{\em e}-folding time (6--19 keV)      & $16.0\pm 2.2$ s  \\
peak intensity (2--19 keV)            & $710\pm 32$ mCrab  \\
blackbody {\em k}T                    & $1.64^{+0.10}_{-0.09}$ keV \\
$R_{\rm km}/d_{10~{\rm kpc}}$         & $9.2^{+1.5}_{-0.9}$ \\
Reduced $\chi^{2~(a)}$                & 1.00  \\
steady emission $^{(b)}$              & $< 6$ mCrab $^{(c)}$ \\
\noalign{\smallskip}
\hline\noalign{\smallskip}
\noalign{$^{(a)}$ 26 d.o.f.; 
 $^{(b)}$ MJD 51452.05--51453.21, $3\sigma$ upper limit, 2--28 keV; 
 $^{(c)}$ In the 2--28 keV band, 1 mCrab $= 3.17\times 10^{-8}{\rm erg~cm}^{-2}{\rm s}^{-1}$.} 
\end{tabular}
\end{flushleft}
\end{table}

\section{SAX~J1752.3$-$3138: Observation and Data Analysis}

The {\em BeppoSAX}-WFCs are two identical coded aperture multi-wire 
proportional counters, each covering a $40^{\circ} \times 40^{\circ}$ field of view, the largest 
ever flown for an arcminute ($1^{\prime}-3^{\prime}$, 99\% confidence) location accuracy X-ray 
telescope. 
The cameras point at opposite directions and operate in the 2--28 keV bandpass. 
The time resolution of the imaging operative mode is 0.488 ms, and the energy resolution is 
18\% at 6 keV.
The imaging capability and the good instrument sensitivity (5-10 mCrab in $10^{4}$ s) allow 
an accurate monitoring of complex sky regions, like the Galactic bulge.

Thanks to their scientific capabilities, the {\em BeppoSAX}-WFCs are well designed to 
systematically investigate transient phenomena lasting from seconds to minutes, like the X-ray 
bursts.
The data of the two cameras is searched for bursts and flares by analysing the time profiles 
of the detectors 
with a time resolution down to 1 s.  
Reconstructed sky images are generated for any statistically significant event, in order to
identify possible new bursters. 
The accuracy of the reconstructed position, which of course depends on the burst intensity, 
is typically better than $5\arcmin$.
This analysis procedure demonstrated its effectiveness throughout the Galactic bulge
WFC monitoring program, leading to the identification of more than $2.2\times 10^{3}$ X-ray 
bursts (or $\sim 1.6\times 10^{3}$ when excluding the Rapid Burster and the Bursting Pulsar)
from 38 different sources (e.g. \cite{Zand00}).

    \begin{figure}[hb]
      \psfig{figure=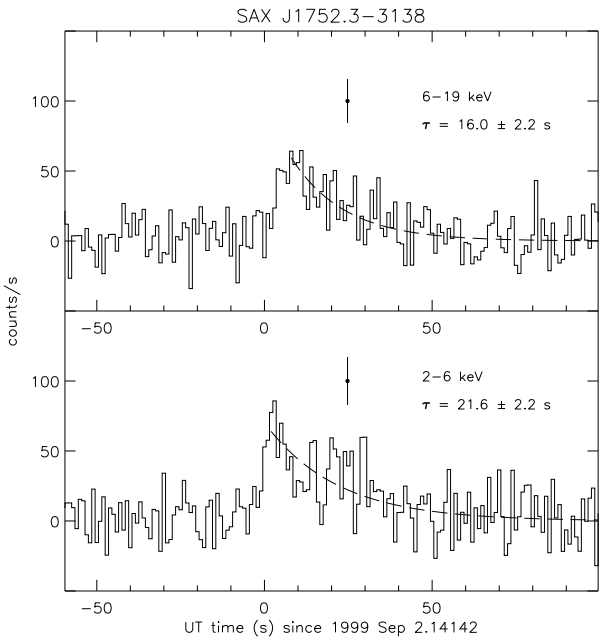,width=8.7cm,clip=t} 
      \caption{
               Energy resolved time histories of the burst.
               The dashed curves are exponential decay fits to the profiles.
	       The typical error bars are also shown.
              }
         \label{Fig1}
    \end{figure}

The X-ray burst from the new source SAX J1752.3$-$3138 was found 
during a post-facto analysis of a 50 ks observation of the Galactic centre region performed 
by the WFC unit 2 on September 2, 1999 (\cite{Cocc99a}).  
The bursting source is located $\sim 3\degr$ SW of Sgr~A,
at $\alpha=17^{\rm h}52^{\rm m}24^{\rm s}$, $\delta=-31^{\circ}37.7^{\prime}$ 
(J2000, error radius $2.9^{\prime}$, 99\% confidence).
The burst peak intensity, corrected for dead time effects, was $\sim 0.7$ Crab.
No steady emission was detected during the whole observation, with a $3\sigma$ upper limit 
of 6 mCrab in the 2--28 keV band.  No other X-ray bursts were reported from the same 
sky position during the observation, even if other events could be missed due to the $\sim 53\%$ 
source covering efficiency associated to the low Earth orbit of {\em BeppoSAX}.
Search for bursts and/or persistent emission from SAX J1752.3$-$3138 was performed on all the data 
available from the 1996-1999 {\em BeppoSAX}-WFC Galactic Bulge monitoring campaigns but, as 
mentioned above, we report no other significant detections. The typical WFC sensitivity (on-axis) 
to burst events is approximately 180 mCrab (2--28 keV), corresponding to a peak of 
$\sim 7\times 10^{37} {\rm erg~s}^{-1}$ ($\sim 30\%~{\rm L}_{\rm Edd}$) for a 10 kpc distance. 
This value is given for a $6 \sigma$ detection of a burst with exponential decay and 
characteristic time of 10 s.

In Fig. 1 the time profiles of the burst, obtained in two energy bands, are shown. 
The time histories of the bursts are constructed by accumulating only 
the detector counts associated with the shadowgram of the analysed source, thus improving 
the signal-to-noise ratio of the profile.  
The background, which is the sum of (part of) the diffuse X-ray background, the particles 
background and the contamination of other sources in the field of view, has been subtracted.
Source contamination is the dominating background component for crowded sky fields like 
the Galactic bulge.  Nevertheless, the probability of source confusion during a short 
time-scale event (10--100 s) like an X-ray burst is negligible.

The burst spectrum of SAX J1752.3$-$3138 is consistent with absorbed blackbody 
radiation with average colour temperature of $\sim 1.6$ keV (see Table 1). 
Due to the rather poor statistics, the $N_{\rm H}$ parameter could not be satisfactorily 
constrained, so it was kept fixed according to the interpolated value for the source sky 
direction, $5.64\times 10^{21} {\rm cm}^{-2}$ (\cite{Dick90}).
An absorbed power-law model does not fit the data adequately.

Time-resolved spectra were accumulated for the observed burst, in order to study the time
evolution of the spectral parameters (Table 2).  The spectra are all consistent with blackbody 
radiation, and spectral softening is observed during the burst decay, in agreement with 
the longer decay time associated to the low-energy time profile (see Fig. 1).
Blackbody spectra allow to determine the relationship between the average radius of the 
emitting sphere $R_{\rm km}$ (in units of km) and the source distance $d_{\rm 10~kpc}$ 
(in units of 10 kpc).
In Fig. 2 the time histories of the measured $R_{\rm km}$/$d_{\rm 10~kpc}$ ratios 
are shown, assuming isotropic emission and not correcting for gravitational redshift and 
conversion to true blackbody temperature from colour temperature (see \cite{Lewi93} 
for details).

    \begin{figure}[ht]
      \psfig{figure=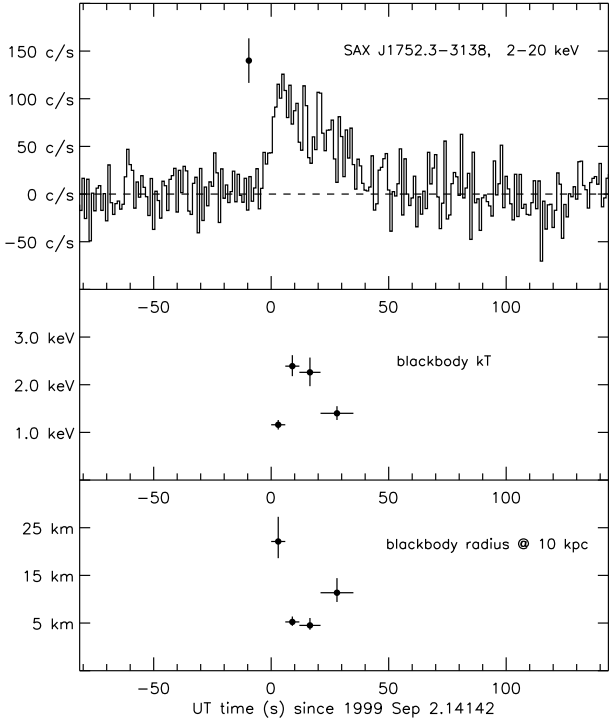,width=8.7cm,clip=t}
      \caption{
               Time history of the burst, the blackbody colour temperature and the 
               blackbody radius, from upper to lower panel, respectively.
              }
         \label{Fig2}
   \end{figure}

\begin{table}[hb]
\caption{ Time resolved spectral analysis of the burst
}
\protect\label{t:hc}
\begin{flushleft}
\begin{tabular}{lccc}
\hline
\hline\noalign{\smallskip} 
time range$^{(a)}$ & {\em k}T (keV) & $R_{\rm km}/d_{10~{\rm kpc}}$  & $\chi^{2~(b)}_{\rm r}$ \\ 
\noalign{\smallskip} \hline\noalign{\smallskip} 
 $0 - 6$ s      & $1.16\pm 0.10$         & $22.1^{+5.2}_{-3.5}$ & 0.73  \\
 $6 - 12$ s     & $2.39^{+0.23}_{-0.21}$ & $5.2^{+1.1}_{-0.8}$  & 0.65  \\
 $12 - 21$ s    & $2.26^{+0.31}_{-0.29}$ & $4.5^{+1.6}_{-0.9}$  & 1.10  \\
 $21 - 34$ s    & $1.40^{+0.15}_{-0.14}$ & $11.4^{+3.1}_{-1.9}$ & 0.71  \\
\noalign{\smallskip} \hline\noalign{\smallskip}
\noalign{ $^{(a)}$ Time since burst start (see Table 1);
          $^{(b)}$ 27 d.o.f.}
\end{tabular}
\end{flushleft}
\end{table}

\section{Discussion}

\subsection{The neutron star LMXB SAX~J1752.3$-$3138}
As anticipated, the spectral and timing properties of the burst detected 
from SAX J1752.3$-$3138 are clearly suggestive of a type I event (\cite{Hoff78}), typically associated to 
weakly magnetised neutron stars in low-mass binary systems (e.g. \cite{Lewi93}). 
In fact the blackbody emission and the measured colour temperatures of $\sim 1.6$ keV are consistent with 
this hypothesis. Spectral softening is observed in the time resolved spectra of the bursts (Table 2), 
and the bursts time profiles can be fitted with exponential decays whose 
characteristic times are energy dependent, being shorter at higher energies (Fig. 1). 

The time history of the blackbody radius as obtained from the time resolved spectral analysis of 
the burst (Fig. 2, lower panel) shows evidence for radius expansion during the very first 
seconds of the event. This is generally interpreted as adiabatic expansion of the neutron star photosphere
during a high luminosity (Eddington-limited) type I burst. The WFC lightcurve of this event 
is not sensitive enough to show evident flat-top or double-peaked profile (e.g. \cite{Lewi93}), 
due the prompt photospheric expansion (see later).
An expansion by a factor $\ga 3$ with respect to the average decay radius is observed, while lower 
luminosity bursts show almost constant blackbody radius during the whole event. The observed 
expansion is consistent with the time profiles in Fig. 1, as the soft emission precedes by 
$\sim 4$ s the harder one. Relatively high luminosity and no hard emission ({\em k}T is $\sim 1$ keV) 
in the very first seconds of the burst imply high values of the blackbody radius.  If we narrow the time 
bin, concentrating on the first 4 s of the burst, we obtain an even larger radius, $33^{+10}_{-6}$ km 
for a 10 kpc distance. This points to a prompt expansion of the neutron star photospere after the ignition.
The fast rise and the prompt photospheric expansion are suggestive of helium burning, while the duration of the 
burst ({\em e}-folding time of $\sim 20$ s), rather long for a pure helium flash, indicates that 
the burning environment was not hydrogen-free (\cite{Lewi93,Bild00}).

Eddington-luminosity X-ray bursts can lead to an estimate of the source distance, if we 
assume isotropic emission and an Eddington bolometric luminosity of $2.5\times 10^{38} {\rm erg~s}^{-1}$, 
which is appropriate for a $1.4~{\rm M}_{\odot}$ neutron star and helium-rich fuel. 
This value also includes a moderate (20\%) gravitational redshift correction (\cite{VanP94}).
Taking into account the observed peak flux of the burst (Table 1), which extrapolates to an 
unabsorbed bolometric intensity of $(2.48\pm 0.21)\times 10^{-8}{\rm erg~cm}^{-2}{\rm s}^{-1}$, 
we obtain $d=9.2\pm 0.4$ kpc. 
The reported error is $1\sigma$, but the systematic uncertainties due the assumption of standard burst 
parameters are of course larger. \\
Assuming a Crab-like spectrum, we derive for the source persistent (bolometric) luminosity an upper limit of 
$\sim4.6\times 10^{36}{\rm erg~s}^{-1}$ when the burst was observed (Sep.~2, 1999).  
An average upper limit twice as low applies to the whole 1996--2000 data set.
Taking into account the inferred distance of 9.2 kpc, the average blackbody radius during the decay of 
the burst was $\sim 6.7$ km. This value is within the range commonly observed (e.g. \cite{VanP78}; also 
\cite{Cocc00} for recent results).
For a canonical $1.4~{\rm M}_{\odot}$ neutron star with $R = 10$ km, we derive an upper limit of
$4.0\times 10^{-10}{\rm M}_{\odot}{\rm yr}^{-1}$ ($8.3\times 10^{3}{\rm g~cm}^{-2}{\rm s}^{-1}$) to the 
accretion rate.

\subsection{A new class of low-luminosity bursters?}
SAX J1752.3$-$3138 is not the only example of a WFC-discovered X-ray burster with no persistent emission 
observed. In 't Zand et al., 1998, already reported on the discovery of two burst sources (SAX J1753.5$-$2349,
SAX J1806.5$-$2215) with no measured steady emission. Also GRS 1741.9$-$2853 and 1RXS J171824.2$-$402934 did 
not show persistent emission at the time when bursts were discovered (\cite{Cocc99b} and \cite{Kapt00}), even 
though GRS 1741.9$-$2853 was once detected with a luminosity of $\sim 2\times 10^{36} {\rm erg~s}^{-1}$ 
in 1990 (\cite{Suny90}). 
The upper limits on their luminosities not being very constraining (of the order of 
$10^{36}{\rm erg~s}^{-1}$), it is unclear if these sources are to be regarded as weak persistent emitters 
instead of transients.  But, as X-ray bursts are rarely observed (in 4.5 y WFC monitoring, 3 from SAX J1806.5$-$2215 
and GRS 1741.9$-$2853, only 1 from SAX J1752.3$-$3138, SAX J1753.5$-$2349 and 1RXS J171824.2$-$402934), the possibility 
that these sources are persistent emitters with luminosities around or just below $10^{36}{\rm erg~s}^{-1}$ seems 
unlikely.
In fact more regular bursting behaviour is to be expected in this case, as typically observed in the 
medium-luminosity (atoll) NS LMXB (\cite{Lewi93}). \\
So these sources could be usually quiescent systems where occasional mass transfer to the neutron star occurs, thus 
causing type I bursts.  If so, they should not be very different from other weak transients observed by the 
WFCs, where burst activity was observed several days after the outburst (e.g. XTE J1709$-$267, 
SAX J1712.6$-$3739, SAX J1750.8$-$2900, SAX J1810.6$-$2609, SAX J1808.4$-$3658, see e.g. \cite{Zand00}).
The rather large distance (in excess of 8 kpc for SAX J1752.3$-$3138 and GRS 1741.9$-$2853) and the 
incomplete time coverage by sensitive enough all-sky monitors could explain the missed detection of the 
(reasonably weak, a few $10^{36}{\rm erg~s}^{-1}$) outbursts triggering the occasional type I burst 
activity of these sources. \\
There is another intriguing possibility left: this small sample of burst-only objects could be part of a class of 
intrinsically weak persistent bursters (${\rm L} \la 10^{35}{\rm erg~s}^{-1}$), whose observation would be 
very useful to investigate the type I bursting behaviour at low accretion rates. The sporadic burst 
activity of such sources would be then related to their unusual accretion regime instead of their transient 
behaviour.  Actually this could be the case for 1RXS J171824.2$-$402934, whose {\em ROSAT} observations suggest weak persitent 
(even though variable) activity, with bolometric luminosity, taking into account the inferred distance of 
$\sim 7$ kpc (\cite{Kapt00}), not exceeding $\sim 2\times 10^{35}{\rm erg~s}^{-1}$.
 Moreover, an X-ray burst at very low persistent luminosity (${\rm L} \la 10^{33}{\rm erg~s}^{-1}$)
was observed by Gotthelf \& Kulkarni (1997) in the globular cluster M28. 
But that burst was notably subluminous ($L_{\rm peak} \sim 4\times 10^{36}{\rm erg~s}^{-1}$), perhaps due to magnetic 
confinement of the burning fuel (\cite{Gott97}), while the bursts in the WFC sample reached peak luminosities of 
$\sim 10^{38}{\rm erg~s}^{-1}$, commonly observed in NS-LMXBs.
Further measurements with more sensitive instruments could unveil the nature (triansient or weakly 
persistent) of these burst-only sources.

\begin{acknowledgements}
  We thank the staff of the {\em BeppoSAX Science Operation Centre} and {\em Science
  Data Centre} for their help in carrying out and processing the WFC Galactic Centre
  observations. The {\em BeppoSAX} satellite is a joint Italian and Dutch program.
  M.C., A.B., L.N. and P.U. thank Agenzia Spaziale Nazionale ({\em ASI}) for grant support.
  M.C. also thanks M. Federici (IAS staff) for technical support.
\end{acknowledgements}

\end{document}